# Derivation of Electron Spin Relaxation Rate by Electron-Phonon Interaction Using a New Diagram Method

Nam Lyong Kang[1], Sang Don Choi[2]

[1]*Department of Applied Nanoscience, Pusan National University, Miryang 627-706, Republic of Korea* (E-mail: nlkang@pusan.ac.kr)

[2] *Department of Physics, Kyungpook National University, Daegu 702-701, Republic of Korea*

A formula for the electron spin relaxation rate of an electron-phonon system is derived using a new spring-loop diagram method. The result contains the distribution functions for the electrons and phonons properly. Therefore, all the spin flip and conserving processes are explained from a microscopic point of view, and a physical insight of the quantum dynamics of electron spin in a solid is obtained from the diagram. Photons and phonons are classified according to the spin flip and conserving processes. The formula is used to calculate the electron spin relaxation rate ($\gamma$) in GaAs. The temperature ($T$) dependence of the relaxation rate obtained by a comparison with the experimental result reported by Römer et al. is $\gamma \propto T^{1.94}$.

## I. INTRODUCTION

Recently, intensive experimental and theoretical studies have concentrated on the electron spin dynamics of semiconductors from the viewpoint of both physics and applications [1-10]. It is important to understanding the spin relaxation mechanism for realizing useful spintronics devices. The perturbative approach is a popular method for gaining knowledge on the spin dynamics. Despite the considerable interest, however, a detailed theoretical study from a fully microscopic approach is only the first step [11-17]. The Elliot-Yafet (EY) mechanism [18, 19] or the D'yakanov-Perel' (DP) mechanism [20] can contribute to the relaxation rate. For bulk III-V n-type semiconductors, such as GaAs, the DP and EY processes are the dominant spin-relaxation mechanisms at high and low temperatures, respectively [21, 22].

On the other hand, the diagram method is a well-known scheme for calculating the perturbative terms. The standard diagram method is a powerful tool for studying complex interacting systems in solid state physics and can represent the trajectory of particles well in the intermediate states of the scattering processes. On the other hand, the results obtained using the standard diagram contain the Fermi distribution function for the electrons and the Planck distribution function for the phonons in simple additive forms [23-26], which violates the "population criterion" meaning that they should be contained in multiplicative forms [27]. How the Planck and Fermi distribution functions are included in the electron spin relaxation rate is important because the temperature dependence of the relaxation rate is caused by the distribution functions.

In this study, a formula for the electron spin relaxation rate in a system of electrons interacting with phonons is derived using the spring-loop diagram method. The lineshape function included in the electron spin resonance formula is the same as that derived using the Kang-Choi's projection-reduction method [17]. A diagram that can model the quantum dynamics of electron spins in a solid can provide considerable insight. Photons and phonons are classified with different symbols for a diagrammatic interpretation of each case because the spin flip and conserving processes are determined by the electron-phonon interaction and photons. This formula is used to calculate the electron spin relaxation rate in GaAs and propriety of the theory is obtained by a comparison with the experimental result reported by Römer et al [10].

## II. METHOD

When an electromagnetic wave with angular frequency $\omega$ is applied to a system of electrons interacting with phonons, the linear magnetic susceptibility tensor under a static magnetic field $B$ applied along the $z$-axis is given as follows [17]:

$$\chi_{+-}(\omega) = \frac{g^2\mu_B^2}{\Omega_0} \sum_\alpha \frac{(f_{\alpha-} - f_{\alpha+})\gamma_\alpha(\omega)}{[\hbar\omega - g\mu_B B]^2 + [\gamma_\alpha(\omega)]^2} \qquad (1)$$

where $g$ is the electron g-factor, $\mu_B$ is the Bohr magneton, $\Omega_0$ is the volume of the system, and $f_{\alpha-}(f_{\alpha+})$ is the Fermi distribution function for an electron in the state $\alpha$ with spin down (up). The effect of the electron-phonon interaction is included in the line shape function, $\gamma_\alpha(\omega)$, which is derived using a spring-loop diagram method.

The rules for deriving the line shape function by the diagram method are as follows:

**Rule 1)** Two implicit states with up and down spins are induced between the initial ($\alpha -$) and final states ($\alpha +$) by an electron-phonon interaction.

**Rue 2)** An implicit state is connected to the initial (or final) state by a proper interaction coupling factor (C-factor), $C_{\alpha i,\beta j}(q)$, which is represented by a spring (see Table I) and is defined as

$$C_{\alpha i,\beta j}(q) = <\alpha i|H_{\text{ep}}|\beta j> \qquad (2)$$

where $i, j = +$ (up spin) or $-$ (down spin) and the electron-phonon interaction Hamiltonian $H_{\text{ep}}$ depends on the mode of the phonons with the wave vector $q$.

**Rule 3)** Two implicit transition factors (T-factor), $T_\pm(\alpha i, \beta j)$, exist. One forms a clockwise loop, $T_+(\alpha i, \beta j)$, and the other forms a counterclockwise loop, $T_-(\alpha i, \beta j)$ (see Table I). They are defined as follows:

$$T_\pm(\alpha i, \beta j) \equiv G_{\alpha i,\beta j}(\pm\omega_q) P_\pm(\alpha i, \beta j). \qquad (3)$$

Here, the energy denominator factors (G-factor), $G_{\alpha i,\beta j}(\pm\omega_q)$, are defined as follows:

$$G_{\alpha i,\beta j}(\pm\omega_q) \equiv (\hbar\omega + E_{\alpha i} - E_{\beta j} \mp \hbar\omega_q)^{-1} \qquad (4)$$

where $\hbar\omega_q$ is the phonon energy. By $G_{\alpha i,\beta j}(\pm\omega_q)$, the energy conservation is satisfied, i.e. $E_{\beta j} = E_{\alpha i} + \hbar\omega \mp \hbar\omega_q$. The population factors (P-factor), $P_+(\alpha i, \beta j)$ for the clockwise loop and $P_-(\alpha i, \beta j)$ for the counterclockwise loop, are defined as

$$P_+(\alpha i, \beta j) \equiv (1 + N_q) f_{\alpha i}(1 - f_{\beta j}) - N_q f_{\beta j}(1 - f_{\alpha i}) \qquad (5)$$

$$P_-(\alpha i, \beta j) \equiv N_q f_{\alpha i}(1 - f_{\beta j}) - (1 + N_q) f_{\beta j}(1 - f_{\alpha i}) \qquad (6)$$

where $N_q$ is the Planck distribution function for a phonon with energy $\hbar\omega_q$. The red and blue loops denote the spin flip and conserving processes, respectively. In the loops, the right and left arrows (forward and backward processes) denote the photon absorption and emission processes. The solid (upper) and dotted (lower) half circles correspond to the phonon emission and absorption processes, respectively.

**Rule 4)** A C-factor multiplied by a T-factor becomes an element in the line shape function.

**Rule 5)** Finally, summing all the elements, after summing each element over all the phonon wave vectors and implicit states, the line shape function can be obtained.

**TABLE I.** Symbols used in the diagram for deriving the line shape function.

| Symbol | Quantity | Meaning |
|---|---|---|
| $\alpha-$ ——〰〰—— $\beta+$ | $C_{\alpha-,\beta+}(q)$ | Couples the $\alpha-$ state to the $\beta+$ state and absorbs or emits a phonon with wave vector $q$ |
| $\alpha\pm$ ——〰〰—— $\beta\pm$ | $C_{\alpha\pm,\beta\pm}(q)$ | Couples the $\alpha\pm$ state to the $\beta\pm$ state and absorbs or emits a phonon with wave vector $q$ |
| $\alpha-$ ⟲ $\beta+$ | $T_+(\alpha-,\beta+)$ | $(\alpha- \to \beta+)$ transition accompanying the emission of a phonon and the absorption of a photon $-(\beta+ \to \alpha-)$ transition accompanying the absorption of a phonon and the emission of a photon |
| $\alpha-$ ⟲ $\beta+$ | $T_-(\alpha-,\beta+)$ | $(\alpha- \to \beta+)$ transition accompanying the absorption of a phonon and the absorption of a photon $-(\beta+ \to \alpha-)$ transition accompanying the emission of a phonon and the emission of a photon |
| $\alpha\pm$ ⟲ $\beta\pm$ | $T_+(\alpha\pm,\beta\pm)$ | $(\alpha\pm \to \beta\pm)$ transition accompanying the emission of a phonon and the absorption of a photon $-(\beta\pm \to \alpha\pm)$ transition accompanying the absorption of a phonon and the emission of a photon |
| $\alpha\pm$ ⟲ $\beta\pm$ | $T_-(\alpha\pm,\beta\pm)$ | $(\alpha\pm \to \beta\pm)$ transition accompanying the absorption of a phonon and the absorption of a photon $-(\beta\pm \to \alpha\pm)$ transition accompanying the emission of a phonon and the emission of a photon |

In this study, the electron, photon, and phonon are considered. Table II lists the symbols and meanings.

**TABLE II.** Symbols and meanings of the particles considered in this paper.

| Symbol | Meaning |
|---|---|
|  | Electron with down spin |
|  | Electron with up spin |
|  | Photon participating in spin flip processes |
|  | Photon participating in spin conserving processes |
|  | Phonon participating in spin flip processes |
|  | Phonon participating in spin conserving processes |

## III. LINE SHAPE FUNCTION

According to the rules, Fig. 1 shows all possible processes. The diagram for the line shape function has eight implicit transitions (loops) (Table III).

**TABLE III.** Meaning of the eight loops in Fig. 1.

| Loop | Meaning | |
|---|---|---|
| [1] | Clockwise spin flip process | between an initial spin down state ($\alpha-$) |
| [2] | Counterclockwise spin flip process | and an implicit spin up state ($\lambda+$). |
| [3] | Clockwise spin conserving process | between an implicit spin up state ($\lambda+$) |
| [4] | Counterclockwise spin conserving Process | and a final spin up state ($\alpha+$). |
| [5] | Clockwise spin conserving process | between an initial spin down state ($\alpha-$) |
| [6] | Counterclockwise spin conserving Process | and an implicit spin down state ($\lambda-$) |
| [7] | Clockwise spin flip process | between an implicit spin down state ($\lambda-$) |
| [8] | Counterclockwise spin flip process | and a final spin up state ($\alpha+$). |

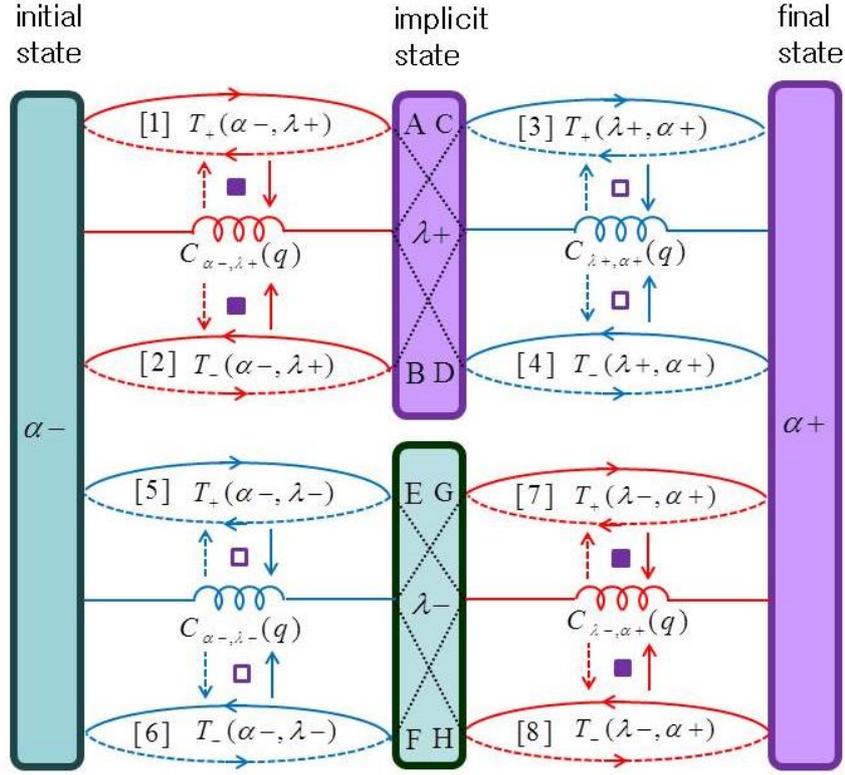

**FIG. 1.** Schematic diagram of all possible electron transition processes. The springs and loops correspond to the interaction coupling factors, $C_{\alpha i,\beta j}(q)$, and the transition factors, $T_{\pm}(\alpha i, \beta j)$, respectively. The A-H processes are the same as those in Eq. (7), e.g., process A consists of the left upper loop [1] and right upper spring.

The implicit states are named as such because they are included only in the line shape function but not in the magnetic susceptibility. Although the implicit transitions are not measured directly, they should be considered in the calculations in the case of an electron system interacting with phonons. The springs show that each implicit state ($\lambda -$ or $\lambda +$) is coupled with an initial spin down state ($\alpha -$) or a final spin up state ($\alpha +$). The red and blue springs show that the initial spin down (final spin up) state is coupled with the implicit spin up (down) state and the initial spin down (final spin up) state is coupled with the implicit spin down (up) state, respectively. Table IV summarizes the meaning of Fig. 1.

TABLE IV. Eight implicit transitions in Fig. 1.

| Processes | Meaning |
|---|---|
| The red vertical solid arrows | Phonons emitted from the electrons during the solid red forward (backward) transitions are absorbed by the red springs. |
| The red vertical dotted arrows | Phonons emitted from the red springs are absorbed by the electrons during the dotted red backward (forward) transitions. |
| The blue vertical solid arrows | Phonons emitted from the electrons during the solid blue forward (backward) transitions are absorbed by the blue springs |
| The blue vertical dotted arrows | Phonons emitted from the blue springs are absorbed by the electrons during the dotted blue backward (forward) transitions. |

From Fig. 1 and the rules, the line shape function can be obtained as follows:

$$\gamma_\alpha(\omega)(f_{\alpha-} - f_{\alpha+})$$
$$= \sum_{q,\gamma} [T_+(\alpha-, \lambda+) + T_-(\alpha-, \lambda+)] |C_{\lambda+,\alpha+}(q)|^2$$
$$+ \sum_{q,\gamma} C_{\alpha-,\lambda+}(q)^2 [T_+(\lambda+, \alpha+) + T_-(\lambda+, \alpha+)]$$
$$+ \sum_{q,\gamma} [T_+(\alpha-, \lambda-) + T_-(\alpha - \lambda-)] C_{\lambda-,\alpha+}(q)^2$$
$$+ \sum_{q,\gamma} |C_{\alpha-,\lambda-}(q)|^2 [T_+(\lambda-, \alpha+) + T_-(\lambda-, \alpha+)]$$
$$= A + B + C + D + E + F + G + H. \qquad (7)$$

Here, $A - H$ are the same as those in Fig. 1, $f_{\alpha-} - f_{\alpha+}$ means the transition from the initial state $(\alpha -)$ to the final state $(\alpha +)$. Eq. (7) is the same as the results obtained using the Kang-Choi's projection-reduction method [17].

The first term (A) in Eq. (7) or Fig. 1 can be interpreted as follows: Note that $T_+(\alpha-, \lambda +) = G_{\alpha-,\lambda+}(+\omega_q) P_+(\alpha-, \lambda +)$. $P(\alpha-, \lambda+)$ means an implicit transition between the initial spin down state $(\alpha -)$ and the implicit spin up state $(\lambda +)$. A phonon (filled purple square) is emitted to the spring (absorbed by an electron) during the solid (dotted) half circle process. The transition forms a loop because the phonon emission process maintains balance with the absorption process. The implicit spin up state is coupled with the final spin up state $(\alpha +)$ by $|C_{\lambda+,\alpha+}(q)|^2$. $G_{\alpha-,\lambda+}(+\omega_q)$ means that the energy of the implicit spin up state is determined by the energies of the initial spin down state, photon, and phonon. Accordingly, an electron undergoes a spin flip implicit transition with phonon emission (or absorption) from the initial spin down state to the implicit spin up state (or vice versa), which is coupled with the final spin up state.

The second term (B) corresponds to the counterclockwise loop between the initial spin down state ($\alpha-$) and the implicit spin up state ($\lambda+$). In the third term (C), the initial spin down state is coupled with the implicit spin up state and a spin conserving implicit transition occurs between the implicit spin up state and the final spin up state ($\alpha+$), where a phonon symbolized by the empty purple square plays a role. The other terms can be interpreted in a similar manner.

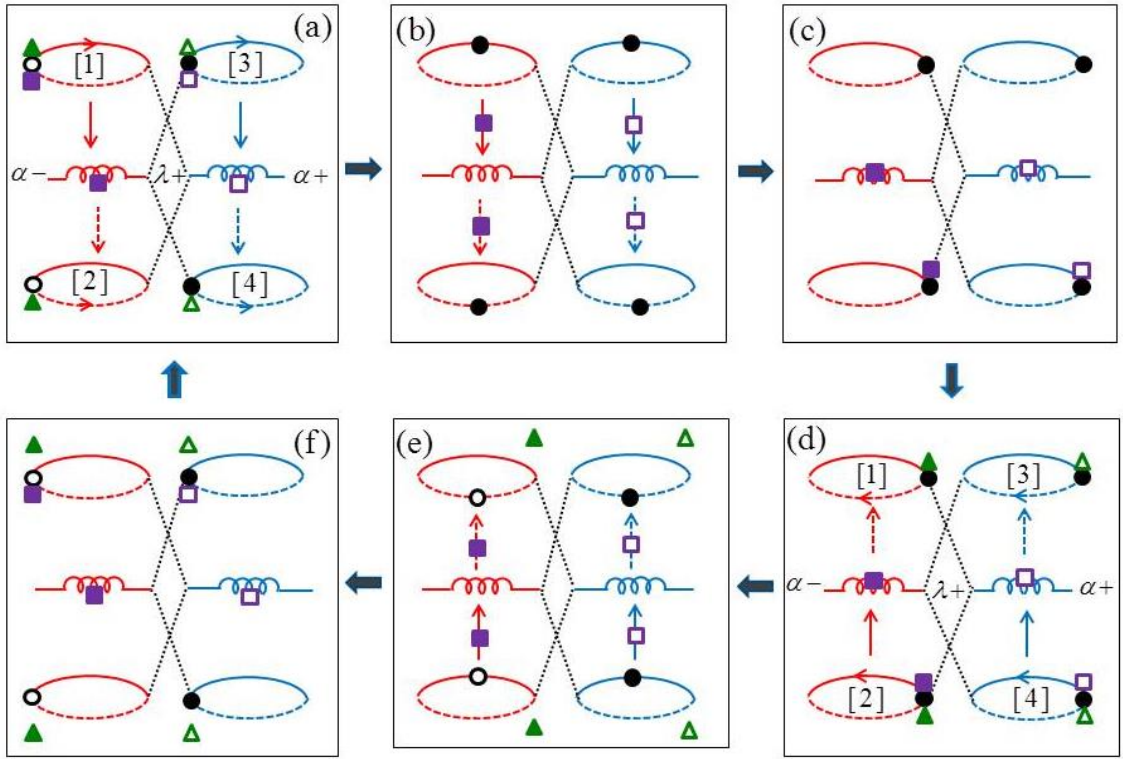

**FIG. 2.** Net implicit transitions of the first four terms in Eq. (7) or Fig. 1. Loops (1), (2), (3), and (4) correspond to the A, B, C, and D processes in Fig. 1 or Eq. (7), respectively. The green triangles, black circles, and purple squares denote the photons, electrons, and phonons, respectively. The filled and empty symbols participate in the spin flip and conserving transitions, respectively.

Figure 2 shows the net implicit transitions of the first four terms in Eq. (7) or Fig. 1. The loop [1], [2], [3], and [4] correspond to the A, B, C, and D processes in Fig. 1 or Eq. (7), respectively. The meaning of the forward process from stage (a) to stage (c) is as follows: (i) An electron (empty black circle) in the initial spin down state of loop [1] emits a phonon (filled purple square) to the spring and absorbs a photon (filled green triangle); (ii) An electron in the initial spin down state of loop [2] absorbs a phonon from the spring and absorbs a photon; (iii) An electron (filled black circle) in the implicit spin up state of loop [3] emits a phonon (empty purple square) to the spring and absorbs a photon (empty green triangle); and (iv) An electron in the implicit spin up state of loop [4] absorbs a phonon from the spring and absorbs a photon. From a state (d) to a stage (f) is a reverse process.

Fig. 3 shows the net implicit transitions of the last four terms in Eq. (7) or Fig. 1. The loops [5], [6], [7], and [8] correspond to the E, F, G, and H processes in Fig. (1) or Eq. (7), respectively.

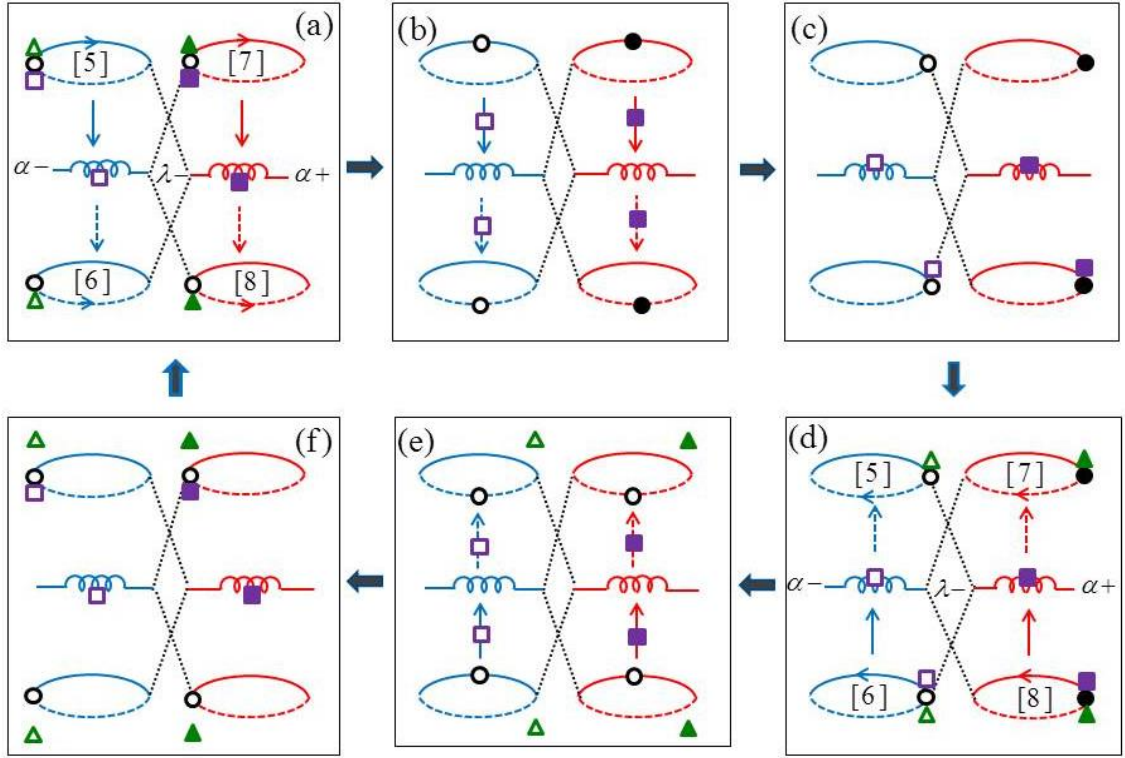

**FIG. 3.** Net implicit transitions of the last four terms in Eq. (7) or Fig. 1. The loops [5], [6], [7], and [8] correspond to the E, F, G, and H processes in the Fig. 1 or Eq. (7), respectively. The symbols are the same as those in Fig. 2.

## IV. APPLICATION

To confirm the correctness of the method, the electron spin relaxation rate ($\gamma$) in GaAs is calculated using Eqs. (1) and (7). If the Lorentzian approximation is assumed for weak scattering, the relaxation rate is related to the line shape function as $\gamma = 2\text{Im}\{\gamma_\alpha(\omega)\}/\hbar$, where Im denotes the "imaginary part". A system of electrons interacting with phonons through the phonon-modulated spin-orbit interaction is considered, for which the interaction Hamiltonian is given as [16, 19]

$$H_{\text{ep}} = \frac{\hbar}{4m^2c^2}[\nabla V_{\text{ep}} \times (\mathbf{p} + e\mathbf{A})] \cdot \sigma \tag{8}$$

where $c$ is the speed of light, $\mathbf{p}$ is the momentum operator of an electron with the effective mass $m$, $\mathbf{A}$ is the vector potential, $V_{\text{ep}}$ is the electron-phonon interaction potential, and $\sigma$ is the Pauli spin matrix. The piezoelectric phonon scattering is one of the main scattering mechanisms in III-V compounds, such as GaAs. The acoustic strain induced by pressure in these materials gives rise to a macroscopic electric field, which is assumed to be proportional to the derivative of the atomic displacement given by

$$\mathbf{u} = \sum_q \sqrt{\frac{\hbar}{2\rho_m \Omega_0 \omega_q}} (b_q + b_{-q}^+) e^{i\mathbf{q}\cdot\mathbf{r}} \hat{\mathbf{e}}_q \tag{9}$$

where $\rho_m$ is the mass density, $b_q^+(b_q)$ is the creation (annihilation) operator for a phonon with a wave vector $\mathbf{q}$, and $\hat{\mathbf{e}}_q$ is the polarization vector. Eq. (8) then becomes

$$H_{\text{ep}} = \sum_{j=x,y,z,\ \alpha,\beta,\ s_\alpha,s_\beta,\ q} [l_j(q)]_{\alpha\beta} a_{\beta,s_\beta}^+ a_{\alpha,s_\alpha} (b_q + b_{-q}^+)(\chi_{s_\alpha} \sigma_j \chi_{s_\beta}). \tag{10}$$

Here, $(X)_{\alpha\beta} \equiv <\alpha|X|\beta>$, $a_{\alpha,s_\alpha}^+(a_{\alpha,s_\alpha})$ is the creation (annihilation) operator for an electron in the state $|\alpha, s_\alpha>$ with spin $s_\alpha$, $\chi_{s_\alpha}$ is the spinor, and

$$\mathbf{l}(q) = \frac{\hbar D_q}{4m^2c^2}[\nabla e^{i\mathbf{q}\cdot\mathbf{r}} \times (\mathbf{p} + e\mathbf{A})] \tag{11}$$

where

$$D_q = P_{\text{pe}} \frac{q^2}{q^2 + q_d^2} \sqrt{\frac{\hbar}{2\rho_m \Omega_0 \omega_q}}. \tag{12}$$

Here, $q_d = \sqrt{n_e e^2/\epsilon \varepsilon_0 k_B T}$ is the reciprocal of the Debye screening length, where $\epsilon$ is the static dielectric constant and $n_e$ is the number density of electrons. In Eq. (12), the proportionality constant (piezoelectric constant) $P_{\text{pe}}$ is used as a fitting parameter.

Figure 4 shows the temperature dependence of the electron spin relaxation rate by piezoelectric phonon scattering in GaAs for $n_e = 2.7 \times 10^{21}$ m$^{-3}$, and $P_{\text{pe}} = 2.4 \times$ $10^{24}$ eV/m, and $f = 30$ GHz, which is the frequency of the incident electromagnetic wave. The result (solid red line) was fitted to the experimental result reported by Römer et al. [10] adding the electron spin relaxation rate by impurity scattering, which was extracted from the experimental data. Impurity scattering is the dominant scattering mechanism below 30 K and the relaxation rate by it is approximately 135 MHz. The electron spin relaxation rate ($\gamma$) by the piezoelectric phonon scattering increases with increasing temperature ($T$) according to the following: $\gamma \propto T^{1.94}$, which show reasonable agreement with the result reported by Yefet [19]. Note that the relaxation rate is proportional to the inverse of the relaxation time ($T_1$). Yefet presented $T_1 \propto T^{-3.5} \sim T^{-2.5}$ for silicon. The increase can be interpreted as due to increase of the number of phonons with temperature.

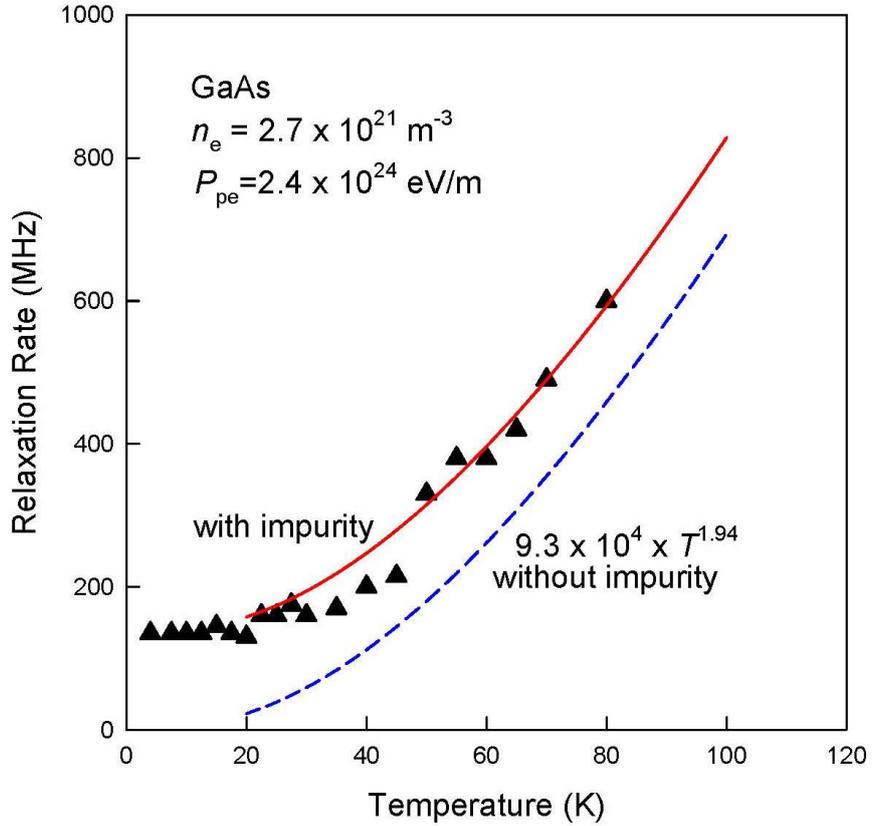

**FIG. 4.** Temperature dependence of the electron spin relaxation rate by the piezoelectric phonon scattering in GaAs for $f = 30$ GHz, $n_e = 2.7 \times 10^{21}$ m$^{-3}$, and $P_{pe} = 2.4 \times 10^{24}$ eV/m. The black triangles present the result reported by Römer et al. [10].

## V. DISCUSSION AND CONCLUSIONS

Relativistic motion of the electron around the nucleus causes the spin–orbit interaction, $H' = -\boldsymbol{\mu}_s \cdot \mathbf{B}_{\text{ob}}$. Here, $\boldsymbol{\mu}_s = g\mu_B \boldsymbol{\sigma}/2$ is the spin magnetic dipole moment of the electron and $\mathbf{B}_{\text{ob}} = \mathbf{E} \times \mathbf{v}/c^2$ is the magnetic field caused by the current due to an electron moving with velocity $\mathbf{v}$ around the nucleus where the electric field $\mathbf{E}$ is related with the electron-phonon interaction potential energy $V_{\text{ep}}$ as $\mathbf{E} = -\boldsymbol{\nabla}V_{\text{ep}}$. The direction of $\boldsymbol{\mu}_s$ is deviated slightly from the $z$-axis because the direction of $\boldsymbol{\nabla}V_{\text{ep}}$, i.e. $\mathbf{B}_{\text{ob}}$ is changed by the electron-phonon interaction. Therefore, when an external static magnetic field, $\mathbf{B} = B\hat{\mathbf{z}}$, and a time-varying magnetic field in the incident electromagnetic wave, $\mathbf{B}_i = B_1 \cos\omega t\, \hat{\mathbf{i}}$ ($i = x, y$), are applied, the electron spin magnetic moments exhibit precession by $\boldsymbol{\mu}_s \times \mathbf{B}$ and their directions can be flipped by $\boldsymbol{\mu}_s \times \mathbf{B}_i$.

The experimental results for the spin relaxation rate could be explained by the formulae developed in this study, and the spin flip and conserving processes could be analyzed properly using the present diagram. Therefore, a proper theory, which is physically acceptable and generally applicable, can be derived using the Kang-Choi's projection-reduction method. The present diagram should be discerned with the Feynman's diagram [28, 29] because the present diagram does not represent the trajectories of the particles in the intermediate stages of the scattering processes. This diagram could be called the "KC spring-loop diagram" because all the contributions to the self-energy terms can be grouped into the topologically-distinct spring-loop diagrams based on the electron-phonon population topology. It is expected that the present method will be applied to other electron transition phenomena.